\documentstyle[12pt,epsf]{article}
\def\mycap#1{ 
\parbox[h]{\textwidth}{\vskip 0.4cm
\footnotesize \baselineskip 0.mm #1 } }

\begin{document}
\begin{center}

{\Large\bf Forward Electron-Phonon Scattering and HTS }\\[1cm]
{\Large
$^{1}${Miodrag L. Kuli\'{c} and Oleg V. Dolgov}$^{2}$}\\[1cm]
{\em
$^{1}$Physikalisches Institut, Uni. Bayreuth, 95440 Bayreuth, Germany\\
$^{2}$Institut f\"{u}r Theoretische Physik, Uni. T\"{u}bingen, Germany 
}\\[2cm]

\parbox[h]{132mm}{
{\bf Abstract.}
Tunneling and point contact spectroscopy show clear phonon features and
together with optic measurements give strong support that the
electron-phonon interaction ($EPI$) is large in $HTS$ oxides. Strong
correlations in $HTS$ oxides renormalize the $EPI$ (and interaction with
impurities) so that the {\bf forward scattering peak (}$FSP${\bf ) }develops
for small hole doping $\delta \ll 1$. The $FSP$ mechanism explains important
properties of the normal and superconducting state.
}

\end{center}

\centerline{\large\bf INTRODUCTION}
\vspace{0.5cm}

The peculiar normal state properties and high $T_{c}$ in $HTS$ oxides can be
explained by the $EPI$ \cite{KulicRev}. In that respect one can rise the
following questions: ${\bf (1)}$ If in some $HTS$ oxides $d-wave$ pairing
and the large $EPI$ are realized, how they are compatible?; $({\bf 2})$ Why
is the pairing in some HTS oxides less anisotropic than in others?; ${\bf (3)}$
Why is the transport coupling constant $\lambda _{tr}(\approx 0.4-0.6)$
(sometimes much) smaller than the pairing coupling constant $\lambda (\sim
2) $? ${\bf (1)}$ - ${\bf (3)}$are explained by strong correlations which
renormalize the $EPI$ giving rise to the forward scattering peak ($FSP$) (at
small $\mid {\bf q\mid }$) in the pairing potential $V({\bf q})$, while the
backward scattering (large$\mid {\bf q\mid }$) is suppressed \cite{Kulic}, 
\cite{KulicRev} - see below.\\

\centerline{\large\bf 
EXPERIMENTAL EVIDENCES FOR THE $EPI$}
\vspace{0.5cm}

{\bf A. Optic and transport measurements. (1) }The dynamical conductivity $%
\sigma (\omega )$ and the resistivity $\rho (T)$ in $HTS$ oxides are
erroneously interpreted by the non-phonon (spin-fluctuation) scattering,
where the analysis is based on the assumption that the {\bf quasiparticle
scattering rate }$\gamma (\omega ,T)=2\mid 
\mathop{\rm Im}%
\Sigma \mid $ and the {\bf transport scattering rate} $\gamma _{tr}(\omega
,T)$ (entering the dynamical conductivity $\sigma (\omega )$) are equal. If $%
\gamma (\omega ,T)\approx \gamma _{tr}(\omega ,T)$ were realized for the $EPI
$ in $HTS$ oxides, then because $\gamma _{tr}^{\exp }(\omega ,T)$ is linear
in $\omega $ up to high $\omega _{s}\sim 2000$ $cm^{-1}$ (where it saturates
with $\omega _{s}\gg \omega _{ph}^{\max }$ - the maximal phonon frequency)
this would make the $EPI$ irrelevant. However, 
this is not the case in $HTS$ oxides,
where the $EPI$ spectral functions $\alpha ^{2}F(\omega )$, $\alpha
_{tr}^{2}F(\omega )$, are spread over the broad $\omega $-range (up to $80$ $%
meV$ - see $Fig.2$ below). This gives that $\gamma ^{EP}(\omega ,T)\neq
\gamma _{tr}^{EP}(\omega ,T)$ \cite{Dolgov}, and that $\gamma ^{EP}(\omega
,T)$ saturates at $\omega \sim \omega _{ph}^{\max }$, while $\gamma
_{tr}^{EP}(\omega ,T)$ saturates at $\omega \gg \omega _{ph}^{\max }$ - see $%
Fig.1$. \cite{Dolgov}. So, the right interpretation of physical quantities
favors the $EPI$ as the origin of the transport properties in the normal
state, including the flatness of the electronic Raman scattering spectra 
\cite{Rashkeev}. The linear resistivity $\rho (T)\sim \lambda _{tr}T$ (for $%
T>T_{c}$) implies a rather small 
$\lambda _{tr}(\ll \lambda \sim 2)\leq 0.4-0.6$. 
This can be explained by the $FSP$ mechanism for the $EPI$ \cite
{Kulic}, \cite{KulicRev} - see below. ({\bf 2}) The Fano shape in the phonon
Raman scattering \cite{PhoRaman} of the normal state and the strongly
superconductivity-induced phonon renormalization of the $A_{1g}$
phonons at $240$ and $390$ $cm^{-1}$ (by $6$ and $18$ $\%$ respectively) in $%
HgBa_{2}Ca_{3}Cu_{4}O_{10+x}$ ($T_{c}=123$ $K$), as well as at $235$ cm$^{-1}
$ and $360$ cm$^{-1}$in $(Cu,C)-1234$ with $T_{c}=117$ $K$ , \ give strong
evidences for the strong $EPI$ in $HTS$ oxide.\\
\newline
\epsfysize=4.1in 
\hspace*{1cm} 
\vspace*{0cm} 
\epsffile{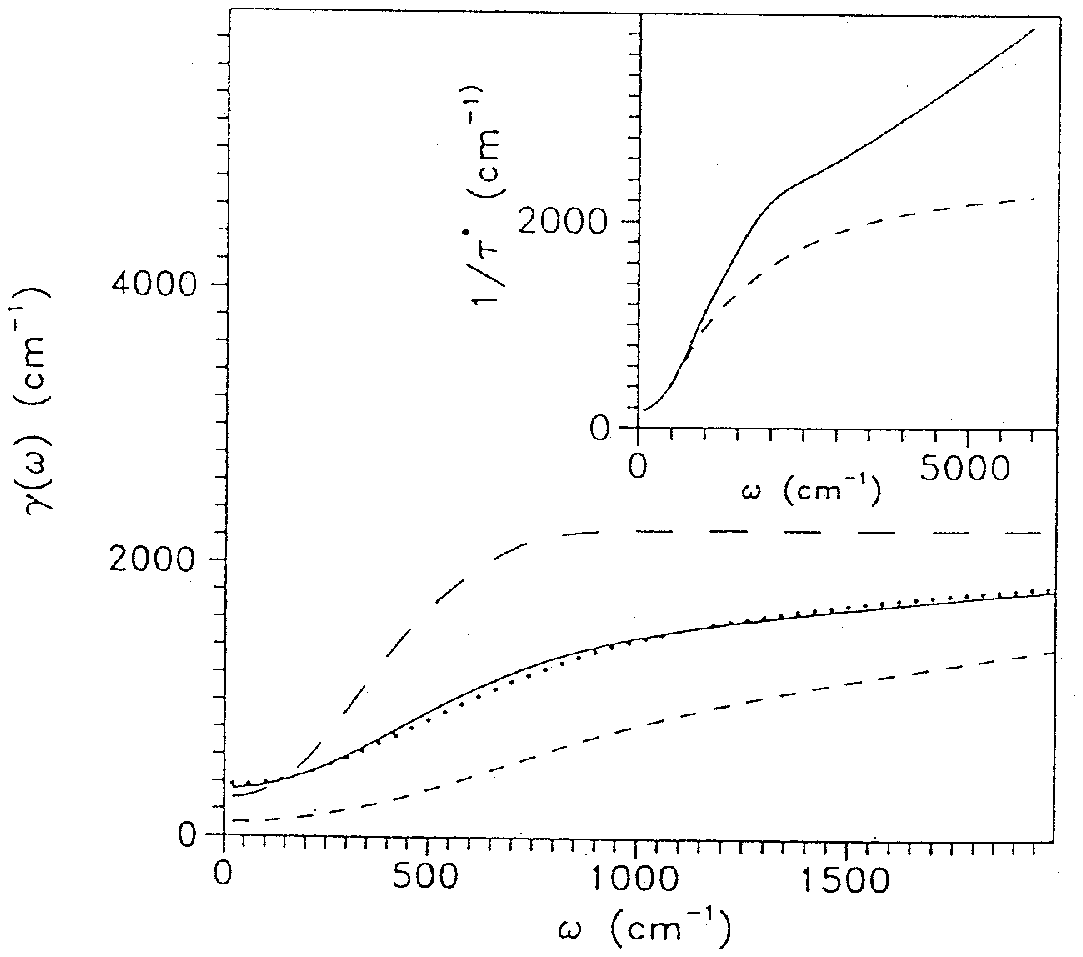}
\newline
\mycap{
{\bf FIGURE~1.} Theoretical $\gamma (\omega ,T)$, $\gamma _{tr}(\omega ,T)$ and $%
\gamma _{tr}^{\ast }(\omega ,T)$ for $\alpha ^{2}F(\omega )$ taken from the
first curve in $Fig.2$ - from \cite{Dolgov}.
}

{\bf B. Tunneling measurements. }The tunneling measurements on $%
Bi_{2}Sr_{2}CaCu_{2}O_{8}$ \cite{Tun2}, \cite{KulicRev} show pronounced
phonon peaks in the extracted $\alpha ^{2}F(\omega )$ for $\omega $ up to $80
$ $meV$ - see $Fig.2$, which coincide with the peaks in the phonon density
of states $F(\omega )$. They show that almost all phonons contribute to the $%
EPI$ giving rise to $\lambda \sim 2$ and $T_{c}\sim 100$ $K$. The pronounced
isotope effect in the under(over) - doped oxides supports the $EPI$
mechanism of pairing \cite{KulicRev} too.\\
\newline
\epsfysize=5.1in 
\hspace*{3cm} 
\vspace*{0cm} 
\epsffile{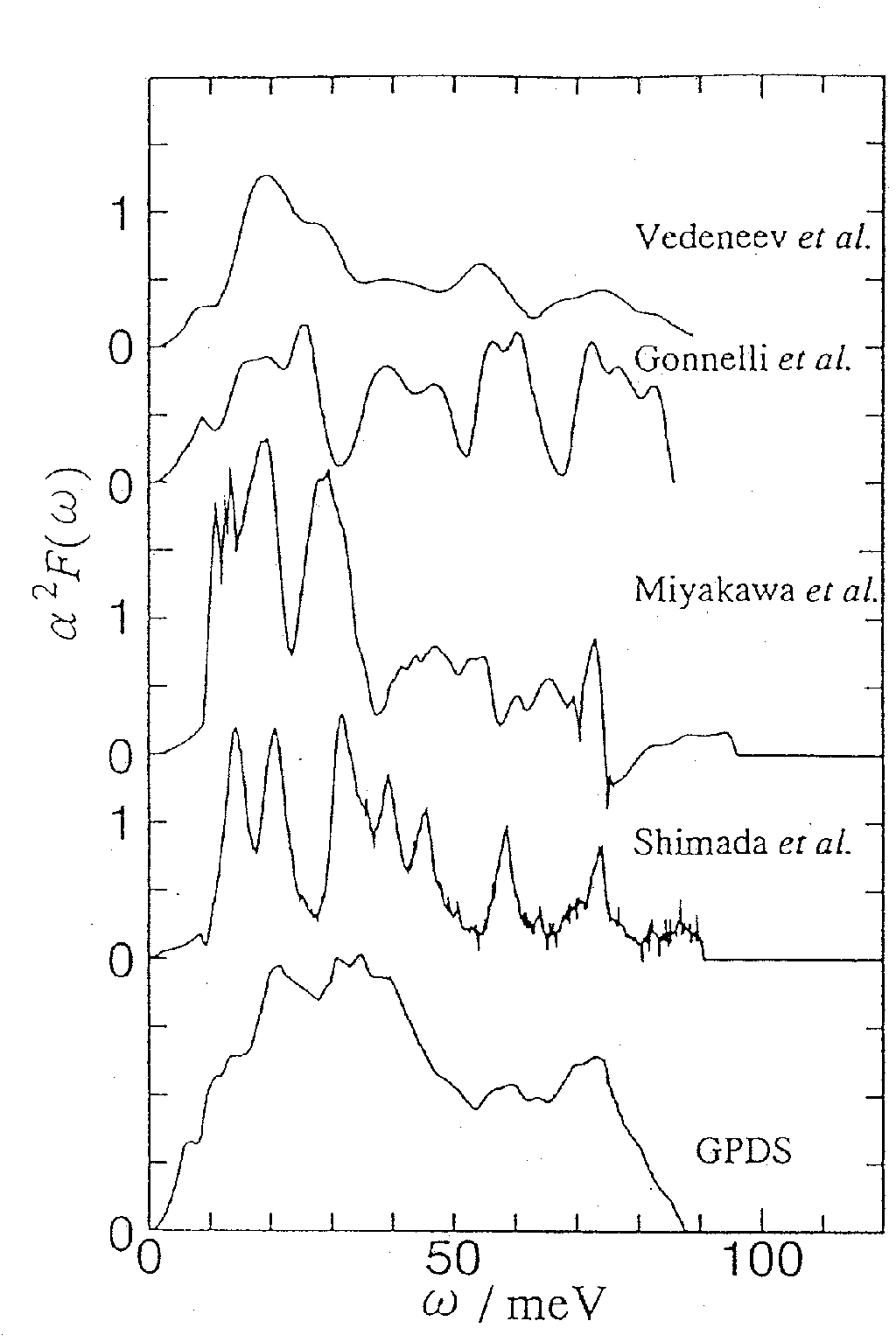}
\newline
\mycap{
{\bf FIGURE~2.} \ $\alpha ^{2}F(\omega )$ from various tunneling measurements. The
GPDS curve is proportional to $F(\omega )$ - from \cite{Tun2}.
}

\centerline{\large\bf 
THEORY OF THE $EPI$ WITH $FSP$\ }
\vspace*{0.5cm}

Several mechanisms produce (separately or combined) the $FSP$ in the $EPI$: $%
(1)$ strong electronic correlations \cite{Kulic}; $(2)$ due to the
ionic-metallic structure of $HTS$ the Madelung energy contribution to the $%
EPI$ is substantial \cite{Abrikosov} and $(3)$ the nearness to the phase
separation \cite{Varelog}. However, $(1)$ in conjuction with $(2)$ is the
most probable mechanism for the $FSP$ in the $EPI$ of $HTS$\ oxides. The
renormalization of the $EPI$ and of the nonmagnetic impurity scattering by
strong correlations is proportional to $\gamma _{c}^{2}({\bf k}_{F},{\bf q})$
(the charge vertex) \cite{Kulic}. At low doping ($\delta $) $\gamma _{c}$ is
peaked at small ${\bf q}$ and suppressed for large ${\bf q}$ \cite{Kulic} - 
{\bf the }$FSP${\bf \ mechanism}, which gives rise to \cite{KulicRev}: $(i)$
the suppression of the $EPI$ in transport properties ($\lambda _{tr}\ll
\lambda $) \cite{Kulic}, i.e. $\rho (T)\sim \lambda _{tr}T$ in the broad
temperature region \cite{Danyl}, and to the non--Drude behavior of $\sigma
(\omega )$ - $Fig.3$; $(ii)$ the pseudogap in the density of states $%
N(\omega )$ - $Fig.4$; $(iii)$ the (normalized) $s-wave$ $EPI$ coupling $%
\Lambda _{1}(\equiv \Lambda _{s})$ and the transport coupling constant $%
\Lambda _{tr}$ decrease by decreasing $\delta $, while the $d-wave$ coupling 
$\Lambda _{3}(\equiv \Lambda _{d})$ increases. By lowering $\delta $ (going
from overdoped to optimally doped systems) the residual Coulomb repulsion
triggers from (anisotropic) $s-$ to $d-wave$ pairing \cite{Kulic}, \cite
{Varelog},\cite{KulicRev} - $Fig.5$; $(iv)$ the robustness of $d-wave$
pairing in the presence of nonmagnetic impurities. In the extreme $FSP$\
limit (with the cutoff $q_{c}\ll k_{F}$ in $V_{ep}({\bf q})$) the equation
for the critical temperature in the weak coupling limit has the form \cite
{Danyl} 
\[
1=V_{ep}T_{c}\sum_{\omega _{n}=-\Omega }^{\Omega }\frac{1}{\omega _{n}^{2}}.
\]
If the phonon frequency $\Omega $ fulfills $\Omega \gg T_{c}$ one obtains $%
T_{c}=\lambda /4N(0)$, where $\lambda =N(0)V_{ep}$. Note, $T_{c}$ $\sim
\lambda $ and large $T_{c}$ is reached even for small $\lambda $ \cite
{Danyl}.\\
\newline
\epsfysize=5.1in 
\hspace*{1cm} 
\vspace*{0cm} 
\epsffile{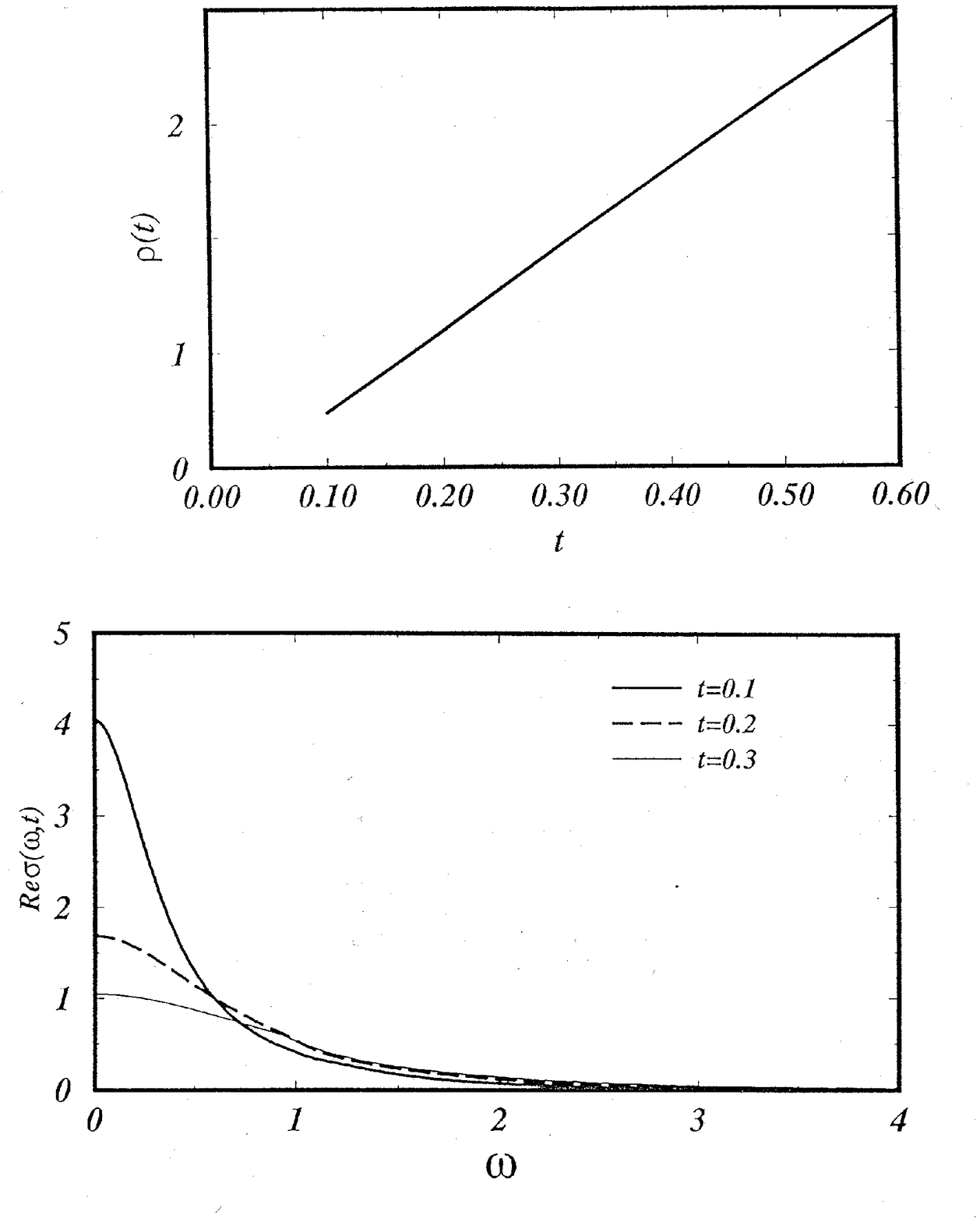}
\newline
\mycap{
{\bf FIGURE~3.}
 \ The prediction of the $FSP$ theory for $\rho (T)$ and $\sigma
(\omega ,T)$ - in arbitrary units; $q_{c}=0$, $t\equiv \pi T/\Omega $, $%
\Omega $ - the phonon frequency; the coupling constant $l(\equiv \lambda
/\pi N(0)\Omega )=0.1$ - from \cite{Danyl}.
}\\
\newline
\epsfysize=4.0in 
\hspace*{1cm} 
\vspace*{0cm} 
\epsffile{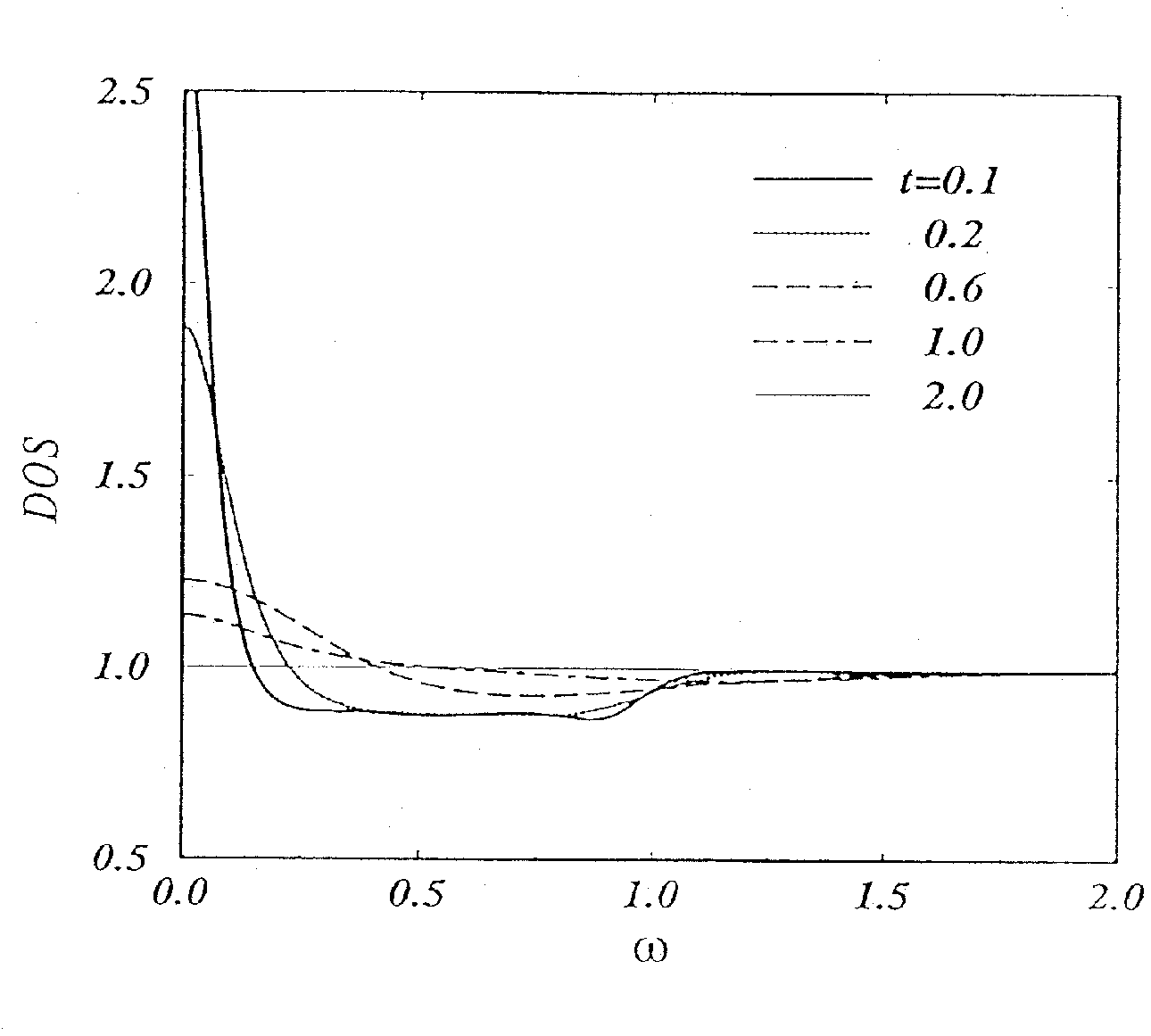}
\newline
\mycap{
{\bf FIGURE~4.} The prediction of the $FSP$ theory for $N(\omega )$ for $q_{c}=0$, $ l=0.1$ and various temperatures $t$ - from \cite{Danyl}.
}\\
\newline
\epsfysize=3.6in 
\hspace*{1cm} 
\vspace*{0cm} 
\epsffile{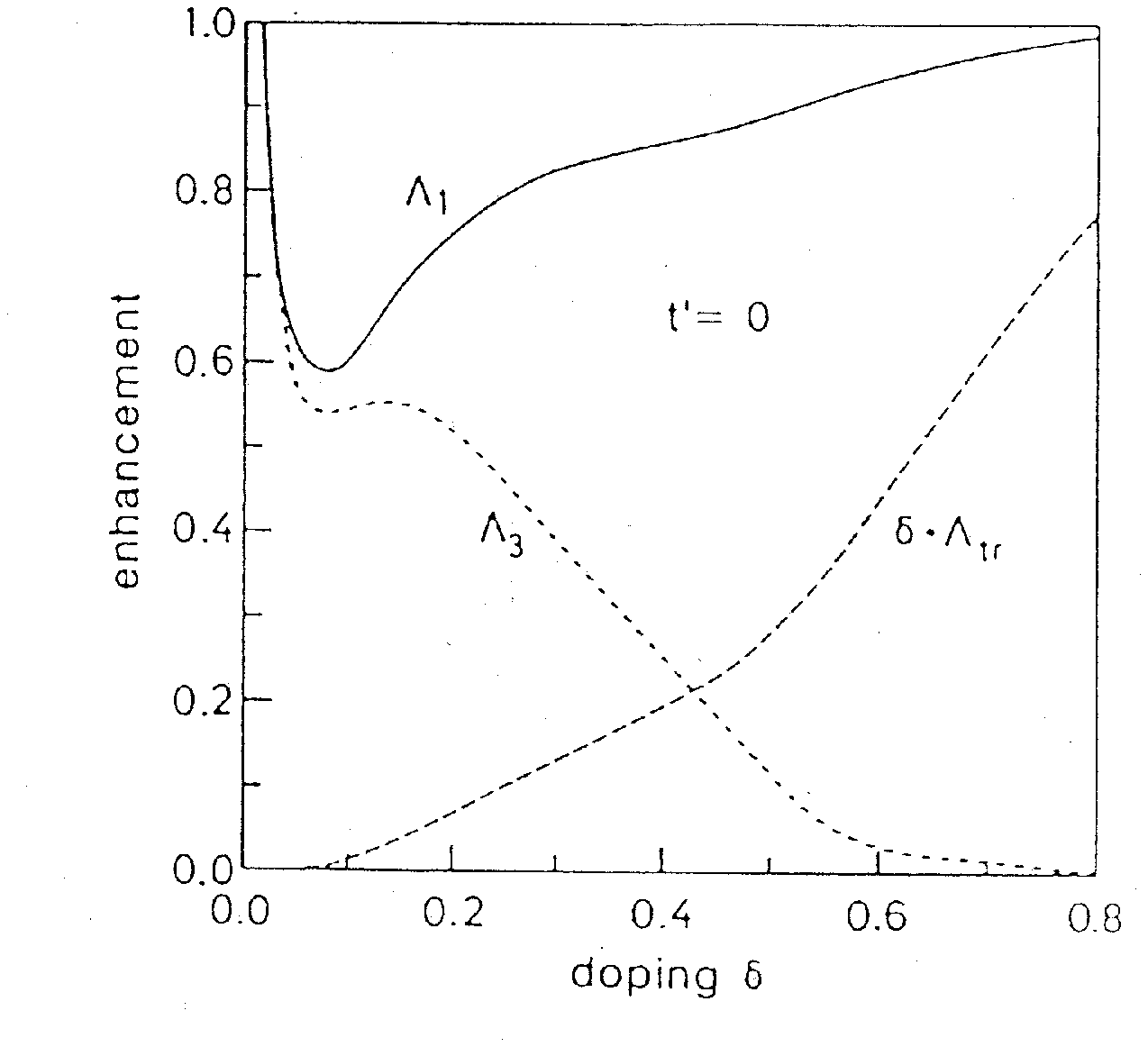}
\newline
\mycap{
{\bf FIGURE 5.} The normalized coupling constants $\Lambda _{1}(\equiv \Lambda _{s})$, $\Lambda _{3}(\equiv \Lambda _{d})$, $\Lambda _{tr}$ as a function of
doping $\delta $ in the $t-t\prime -J$ model ($t\prime =J=0$) - from \cite
{Kulic}.
}

In summary the theory of the $EPI$ with the forward scattering peak can
explain the most relevant properties in the normal and superconducting state
of $HTS$ oxides.\\

\centerline{\large\bf 
ACKNOWLEDGMENTS
}\vspace*{0.5cm}
We are grateful to O.V.~Danilenko and V.~Oudovenko for collaboration on the
$FSP$ subject.

\def\refname{\phantom{xxxxxxxxxxxxx}REFERENCES}

\end{document}